\documentclass{svproc}
\usepackage{xcolor}
\usepackage{amsmath,amssymb}
\usepackage{epsfig}
\usepackage{graphicx,float}
\usepackage{subcaption}
\usepackage{url}
\usepackage[linesnumbered, ruled, vlined]{algorithm2e}
\usepackage{adjustbox}
\usepackage{float}
\usepackage{tikz}
\usepackage{caption} 
\date{}
\begin{document}

\mainmatter

\title{Analysis and predictability of centrality measures in competition networks}

\author{Anthony Bonato\inst{1}\thanks{Supported by an NSERC Discovery Grant.} \and Mariam Walaa\inst{1}}

\institute{Department of Mathematics, Toronto Metropolitan University \\ Toronto, Ontario, Canada}

\maketitle

\begin{abstract}
The Common Out-Neighbor (or CON) score quantifies shared influence through outgoing links in competitive contexts. A dynamic analysis of competition networks reveals the CON score as a powerful predictor of node rankings. Defined in first-order and second-order forms, the CON score captures both direct and indirect competitive interactions, offering a comprehensive metric for evaluating node influence. Using datasets from Survivor, Chess.com, and Dota~2 online gaming competitions, directed competition networks are constructed, and the dynamic CON score is integrated into supervised machine learning models. Empirical results show that the CON score consistently outperforms traditional centrality measures such as PageRank, closeness, and betweenness centrality in classification tasks.

By integrating dynamic centrality measures with machine learning, our proposed methodology accurately predicts outcomes in competition networks. The findings underline the CON score's robustness as a feature in node classification, offering a significant advancement in understanding and analyzing competitive interactions.
\end{abstract}

\section{Introduction}

Centrality measures were used to study real-life networks since the early days of modern network science \cite{bavelas1950,brin1998,freeman1979,katz1953} and, more recently, as predictors in machine learning problems \cite{bucur2020,liu2022,morselli2013}. They are crucial for identifying actors who are influential in a complex network and are often considered to be local metrics. 

Competition networks model adversarial relationships between actors. Examples of competition networks constructed using real-life data include e-sports win networks, and voting networks for reality competition shows. In \cite{bonato2018}, the authors develop a tool to measure the correlation between known winners and their membership in strong alliances within competition networks from reality television shows Survivor and Big Brother. Alliance strength in these competitions is quantified by analyzing the edge densities of known alliances. 

In \cite{bonato2019}, the authors introduce the CON score as a predictor of influential actors in competition networks. More broadly, they proposed the \emph{Dynamic Competition Hypothesis} (or DCH), which suggests that leaders in adversarial networks exhibit a high CON score, high closeness, low in-degree, and high out-degree. In \cite{bonato2022}, the concept of \emph{low-key leaders} was introduced, defined as nodes that maintain influence despite low centrality (from low PageRank). These leaders are identified by contrasting their high CON scores with their PageRank values \cite{brin1998}, demonstrating that influence in competition networks can arise from factors beyond traditional centrality measures.

Our objective in this paper is to analyze the centrality of competition networks constructed from game-based datasets collected from online sources and to build network-based machine learning models capable of accurately predicting known outcomes of these games. We claim that linking centrality measures to adversarial interactions and analyzing them over time provides an effective method for ranking nodes in competition networks, providing a high correlation with ground truth data. We demonstrate that the CON score is a highly effective predictive feature, often outperforming traditional centrality measures such as PageRank, closeness, and betweenness centrality.

The paper is organized as follows. Section~2 introduces the first-order and second-order CON scores and provides definitions of other centrality measures used in this study. Section~3 describes the network datasets and explains the process of feature generation for building machine learning models using the CON score and other centrality measures. Finally, we conclude with a summary of our findings and discuss open problems for future research.

We consider weighted, directed graphs (or \emph{digraphs}) with directed edges in the paper. The \emph{in-neighbors} and \emph{out-neighbors} of node $v$ are denoted by $N^-(v)$ and $N^+(v),$ respectively. The in- and out-degree of $v$ is denoted by $\deg^-(v)$ and $\deg^+(v),$ respectively. Additional background on graph theory may be found in \cite{west}, and more background on complex networks may be found in \cite{bonato2008}.

\section{The CON Score}

High centrality is intuitively one of the key characteristics of an influential node in a competition-based network. A strong player is expected to engage with many others in the network and emerge victorious in these interactions. In this section, we formalize these ideas by defining competition networks and their dynamic extensions.

A \emph{competition network} is a directed graph $G = (V, E)$, where each node $u \in V$ represents a competitor, and a directed edge $(u, v) \in E$ exists if $u$ has defeated $v$ at least once in a competition. This structure captures the competitive interactions and relationships among actors in the network. A \emph{dynamic competition network} extends this concept across multiple rounds or time-steps. It is represented as a sequence of graphs $(G_t: 1 \le t \le T)$, where each $G_t = (V, E_t)$ describes the competition network at time step $t$. The set of nodes $V$ (which represent the competitors) remains fixed across time, while $E_t \subseteq V \times V$ denotes the directed edges (or competitions) occurring at time $t$.

To quantify the competitive interactions between two nodes, we define the \emph{Common Out-Neighbor} (or \emph{CON}) score of $u$ and $v$ as the number of nodes to which both $u$ and $v$ have directed edges. Specifically, $\text{CON}(u, v)$ measures the number of common competitors that $u$ and $v$ have defeated. For a detailed mathematical definition of $\text{CON}(u, v)$, refer to \cite{bonato2019}.

The \emph{CON score of $u$} quantifies the number of common out-neighbors a node shares with the rest of the network. We compute the CON score for every node in a competition network and use it to rank nodes based on competitiveness. Let $v$ be an arbitrary node in graph $G$, and let $\mathbf{A}$ be the adjacency matrix of $G$. To account for first-order neighbors, we consider $\mathbf{A}[i, j]$, the $(i, j)$ entry of $\mathbf{A}$, which represents the number of directed edges from $i$ to $j$, or the number of victories by actor $i$ against actor $j$. This gives:
\[
    \text{CON}_{1}(v) = \sum_{\substack{u, x \in V(G), \\ u \neq v}} \min(\mathbf{A}[v, x], \mathbf{A}[u, x]).
\]

We next introduce a new measure relevant to competition networks: the \emph{second-order CON score}, which considers the number of common out-neighbors up to distance two in the competition network. The purpose of the second-order CON score is to capture competitors of competitors, extending beyond direct competition. For example, if player $u$ competes with player $w$, and player $v$ competes with player $x$, who in turn competes with $w$, the first-order CON score does not capture this $(u, w)$-$(v, w)$ CON pair, but the second-order CON score does. For an illustration of this, see Figure~1.

\begin{figure}[H]
    \centering
    \begin{minipage}{0.3\textwidth}
        \centering
        \begin{tikzpicture}[>={latex}, node distance=2cm]
            \node[circle, draw, fill=white, text=black, minimum size=0.8cm] (u1) at (0, 1) {$u$};
            \node[circle, draw, fill=white, text=black, minimum size=0.8cm] (v1) at (0, 0) {$v$};
            \node[circle, draw, fill=white, text=black, minimum size=0.8cm] (w1) at (1, 0.5) {$w$};
            \draw[->, thick] (u1) -- (w1);
            \draw[->, thick] (v1) -- (w1);
        \end{tikzpicture}\caption*{(a)}
    \end{minipage}
    \hfill
    \begin{minipage}{0.3\textwidth}
        \centering
        \begin{tikzpicture}[>={latex}, node distance=2cm]
            \node[circle, draw, fill=white, text=black, minimum size=0.8cm] (u2) at (0, 1) {$u$};
            \node[circle, draw, fill=white, text=black, minimum size=0.8cm] (v2) at (0, 0) {$v$};
            \node[circle, draw, fill=white, text=black, minimum size=0.8cm] (z2) at (1, 0) {$z$};
            \node[circle, draw, fill=white, text=black, minimum size=0.8cm] (w2) at (2, 0.5) {$w$};
            \draw[->, thick] (u2) -- (w2);
            \draw[->, thick] (v2) -- (z2);
            \draw[->, thick] (z2) -- (w2);
        \end{tikzpicture}\caption*{(b)}
    \end{minipage}
    \hfill
    \begin{minipage}{0.3\textwidth}
        \centering
        \begin{tikzpicture}[>={latex}, node distance=2cm]
            \node[circle, draw, fill=white, text=black, minimum size=0.8cm] (u3) at (0, 1) {$u$};
            \node[circle, draw, fill=white, text=black, minimum size=0.8cm] (v3) at (0, 0) {$v$};
            \node[circle, draw, fill=white, text=black, minimum size=0.8cm] (x3) at (1, 1) {$x$};
            \node[circle, draw, fill=white, text=black, minimum size=0.8cm] (z3) at (1, 0) {$z$};
            \node[circle, draw, fill=white, text=black, minimum size=0.8cm] (w3) at (2, 0.5) {$w$};
            \draw[->, thick] (u3) -- (x3);
            \draw[->, thick] (x3) -- (w3);
            \draw[->, thick] (v3) -- (z3);
            \draw[->, thick] (z3) -- (w3);
        \end{tikzpicture}\caption*{(c)}
    \end{minipage}
    \caption{Three scenarios displaying direct and indirect competition in a network. In (a), both $u$ and $v$ have direct competition with $w$, whereas in (b), only $v$ has indirect competition with $w$, and in (c), both $u$ and $v$ have indirect competition with $w$.}
\end{figure}

To account for second-order neighbors, which correspond to all distance-two paths, we compute the square of the adjacency matrix. Summing both distance-one and distance-two paths gives:
\[
    \mathbf{A}_{2}[i, j] = \mathbf{A}[i, j] + \mathbf{A}^2[i, j].
\]
For the second-order CON score, we have:
\[
    \text{CON}_{2}(v) = \sum_{\substack{u, x \in V(G), \\ u \neq v}} \min(\mathbf{A}_{2}[v, x], \mathbf{A}_{2}[u, x]).
\]

To evaluate the effectiveness of the second-order CON score, we compare it with well-established centrality measures used for analyzing node importance. The \emph{closeness centrality} \cite{bavelas1950} of a node $v$ in $G$ is defined as:
\[
    C(v) = \left( \sum_{u \in V(G) \setminus \{v\}} d(v, u) \right)^{-1},
\]
where $d(v, u)$ is the shortest path distance from $v$ to $u$, if such a path exists. The \emph{betweenness centrality} \cite{freeman1979} is defined as:
\[
B(v) = \sum_{x, y \in V(G) \setminus \{v\}} \frac{\sigma_{xy}(v)}{\sigma_{xy}},
\]
where $\sigma_{xy}(v)$ is the number of shortest directed paths between $x$ and $y$ that pass through $v$, and $\sigma_{xy}$ is the total number of shortest directed paths between $x$ and $y$.

A final, well-known centrality measure we consider is PageRank \cite{brin1998}, albeit applied to the digraph formed by reversing the edges of the network. For more on PageRank, see \cite{bonato2008}.

\section{Experimental Design and Methods}

We consider a variety of game data to construct our competition networks. We start with a dataset for \emph{Survivor}, a famous reality competition show that began in the U.S.\ in 2000. The American \emph{Survivor} competitions are structured consistently across all independent seasons. In each season, approximately 16 players are divided into tribes and must compete to win the show's grand prize. Individuals form latent, shifting alliances within tribes to eliminate weak or threatening tribe members. Despite the team-based competition, each castaway must ultimately fight for their chance at the grand prize. More details about the competition are provided in \cite{bonato2018}. In the \emph{Survivor} network, individual players are represented as nodes, and votes between players form directed edges.

The second dataset we work with is from the \emph{Chess.com Titled Tuesday} competitions. These competitions follow FIDE rules and guidelines as outlined in \cite{fide2025}. Players are initially ranked based on strength (rating), FIDE title, and alphabetically. This ranking, which is referred to as the \emph{Initial Order (or \emph{IO})}, determines the \emph{Pairing Numbers} (or \emph{PN}). After each round, \emph{scoregroups} are formed, consisting of players with the same score. A \emph{pairing bracket} is considered homogeneous if all players within it have the same score; otherwise, it is classified as heterogeneous. Players within a scoregroup may sometimes have different scores because of \emph{downfloaters} or \emph{upfloaters}, who are moved from their original brackets to facilitate pairings. After each round, players are sorted lexicographically, first by scoregroup and then by IO. Pairings occur once scoregroups and pairing brackets are finalized. In general, the FIDE pairing system ensures that players compete with others of similar skill in the first round and are paired in subsequent rounds based on performance and overall skill. In the Chess.com network, players are represented as nodes and directed edges correspond to match outcomes.

Finally, we use a dataset from \emph{Dota 2}, a Multiplayer Online Battle Arena (or MOBA) game developed in 2013. The original game, Dota, was among the first to establish the MOBA genre, becoming widely popular. Dota 2 has since achieved over 80 million accounts in the past decade. The Dota 2 Professional League matches represent a subset of Dota 2 games that are collected, tracked, and updated weekly on Kaggle, providing detailed insights into matches and team performance. The Dota 2 competition network is constructed by representing each team as a node and connecting them with directed edges that indicate match outcomes, where an edge points from the winning team to the losing team.

Our datasets were selected based on availability and relevance. Experiments conducted on these datasets demonstrate that the CON score can serve as a powerful predictive feature in machine learning models for node classification. We construct the competition networks as follows.

\begin{enumerate}
    \item \emph{Survivor}: Voting data for 46 seasons of \emph{Survivor} \cite{survivor2py} was downloaded from the public GitHub repository, \texttt{survivoR2py}. Two files were primarily used to construct the network: \texttt{vote\_history.csv}, which contains castaways and their votes in each episode, and \texttt{castaways.csv}, which contains each castaway's outcome at the end of the season (for example, sole survivor, runner-up, or $n$-th voted out). After data cleaning, a total of 726 castaways from 46 seasons were extracted, with 3,662 tribal council votes across an average of 12 episodes per season.

    \item \emph{Chess.com}: Tournament data from Chess.com's Titled Tuesday webpage \cite{chess.com} was scraped. Two files containing results and pairings for each tournament round were utilized. The tournaments operate in a Swiss format, with players ``dropping in'' weekly to participate in early or late games over 11 rounds. Each Chess.com player has a Glicko rating that determines their overall ranking in terms of chess skill. The Chess.com dataset we considered had 933 nodes and 16,571 edges.

    \item \emph{Dota 2}: Professional league match data was downloaded from Kaggle, and OpenDOTA rankings were retrieved using the OpenDOTA API in Python. Each Dota 2 team has a Glicko rating that determines their rank in terms of Dota 2 skill. Two files were used to construct the network. First, there was \texttt{main\_metadata.csv}, which contains details about each match, including team IDs and whether a team was radiant or dire, and second, \texttt{teams.csv}, which contains team names. The Dota 2 dataset we considered had 493 nodes and 2,413 edges.
\end{enumerate}

\begin{figure}[h!]
    \centering
    \begin{minipage}{0.4\textwidth}
        \centering
        \includegraphics[width=\textwidth]{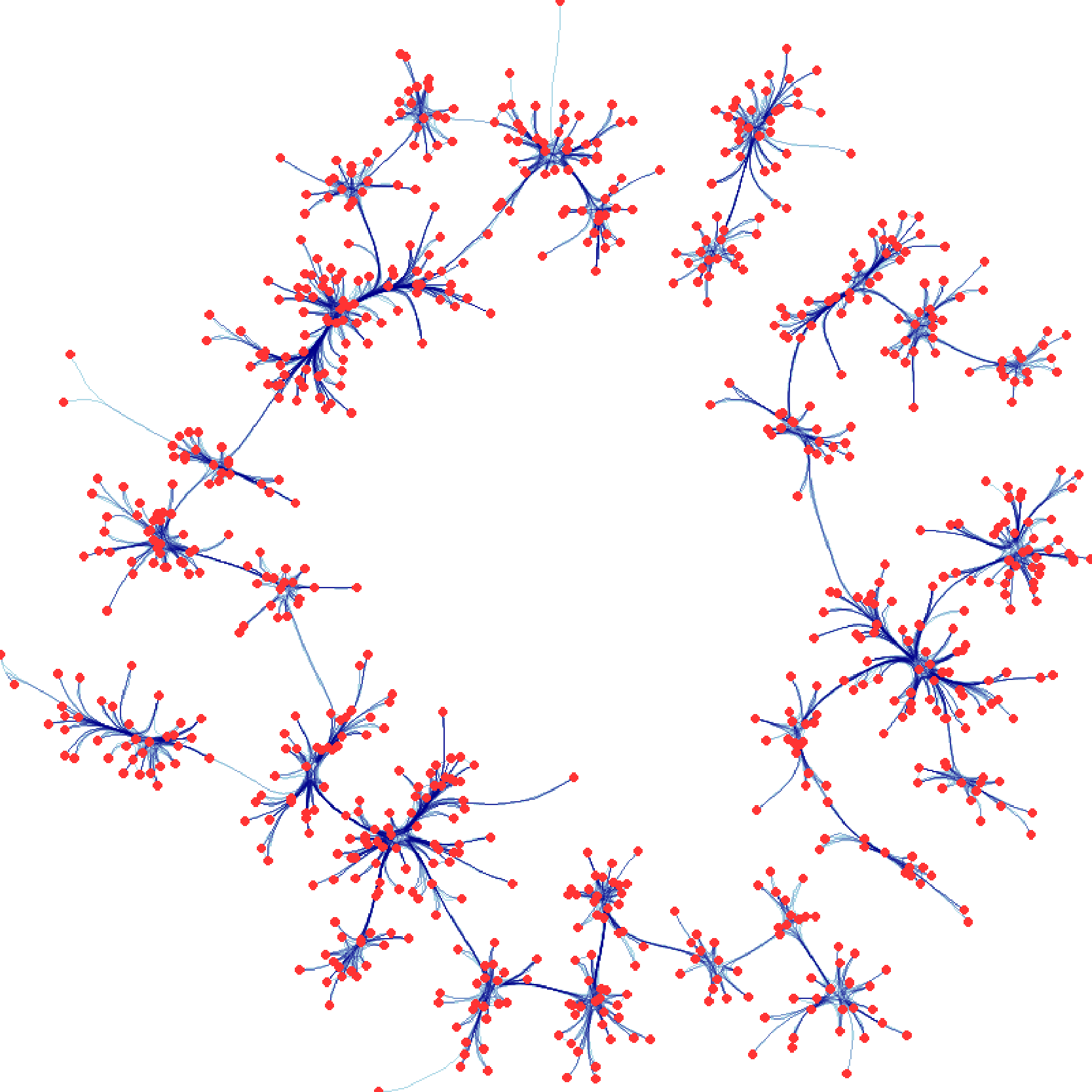}
        \caption*{(a) Survivor}
    \end{minipage}
    \hfill
    \begin{minipage}{0.4\textwidth}
        \centering
        \includegraphics[width=\textwidth]{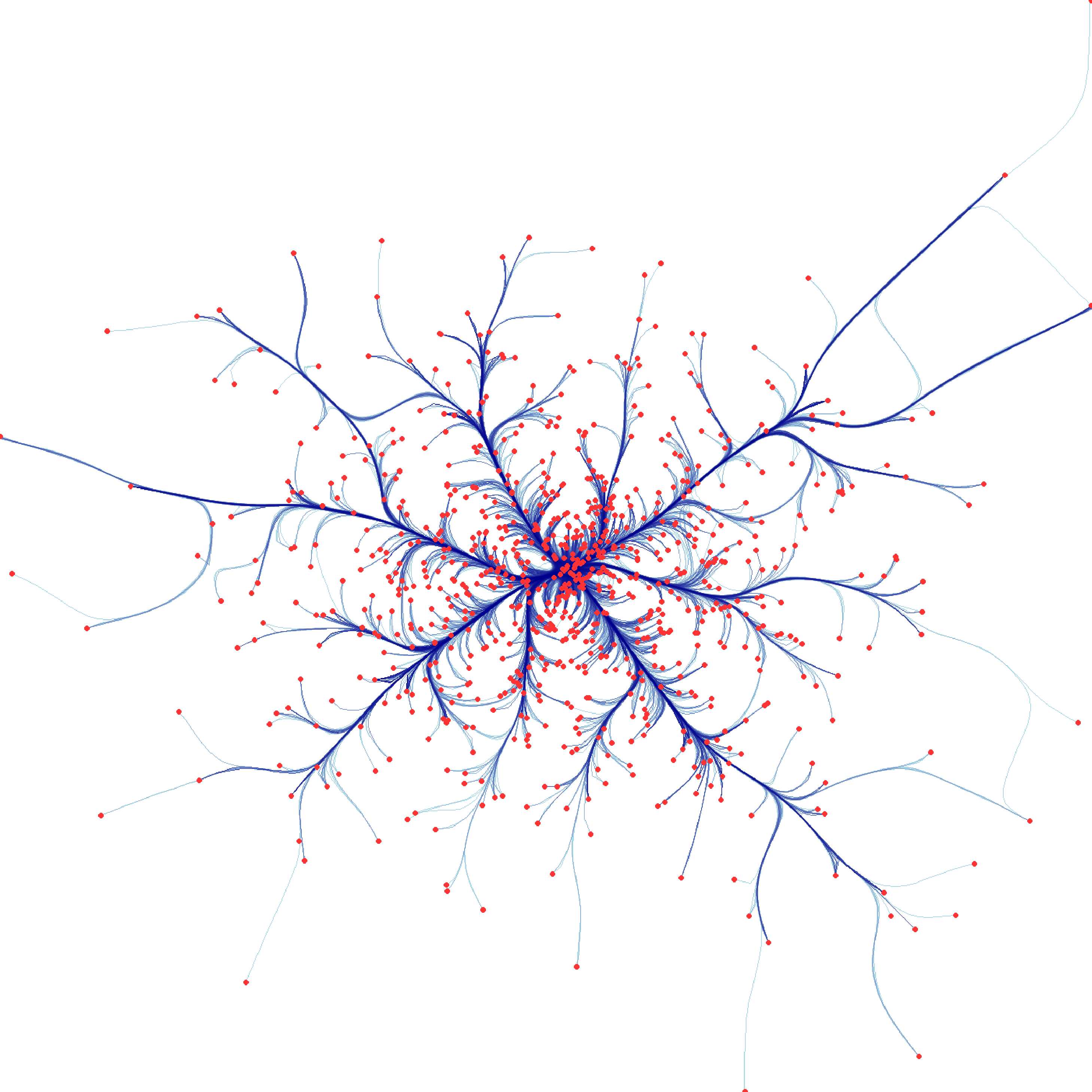}
        \caption*{(b) Chess.com}
    \end{minipage}
    \hfill
    \begin{minipage}{0.4\textwidth}
        \centering
        \includegraphics[width=\textwidth]{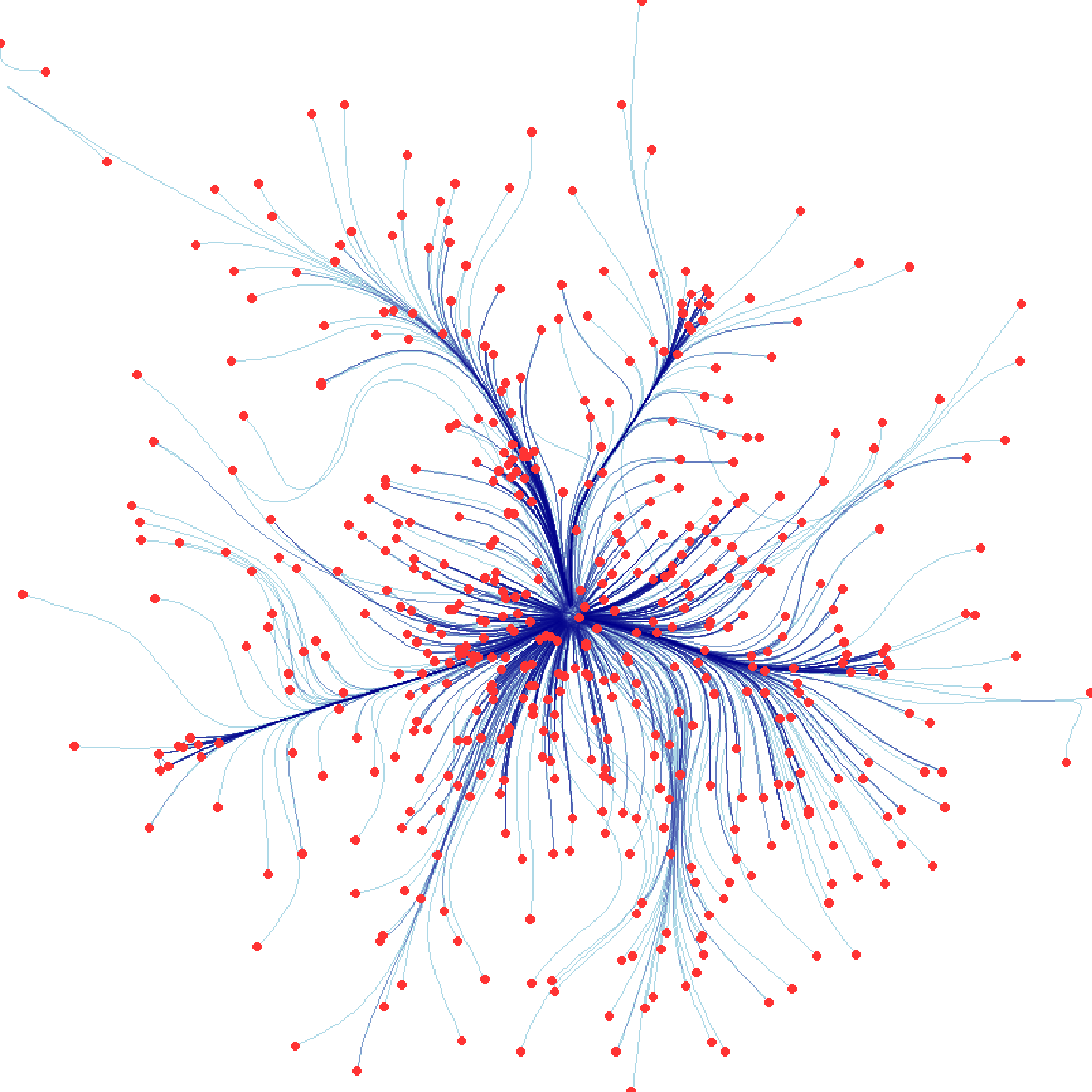}
        \caption*{(c) Dota 2}
    \end{minipage}
    \caption{Force-directed, edge-bundled networks based on game datasets; see \cite{bednar2016}. In (a), smaller disconnected networks organized in a circular structure represent 46 independent seasons of Survivor. In (b), a highly centralized, connected network represents Chess matches between players. In (c), disconnected networks with a dense central core represent top teams playing matches in Dota 2. Both (b) and (c) show low-interaction players/teams on the periphery, with high-interaction players/teams in the core.}
\end{figure}

An important aspect of supervised machine learning modeling is the acquisition of ground truth labels. In our datasets, ground truth labels represent the true ranking of nodes in a competition network (for example, player rankings or outcomes). These labels are used to evaluate how centrality-based predictions align with actual outcomes. Ground truth rankings are compared with the model’s predictions based on centrality measures, allowing for the approximation of node influence. True rankings are determined by accurate, up-to-date Glicko ratings or holistic game outcomes and are not specific to the sample of game data collected. At the end of each season, each Survivor participant has an \emph{outcome} describing whether they are the sole survivor, a runner-up, or the $n$-th voted out. Similarly, each Chess.com player has a Glicko rating that defines their rank in terms of chess skill, and each Dota 2 team has a Glicko rating that defines their rank in terms of Dota 2 skill.

As shown in Figure~2, the nature of the competition networks resulting from these games varies, as different factors across the three datasets drive the presence of an edge between two actors. For instance, in the Chess.com data, players are paired with those of similar skill and game outcomes, while in Survivor, players vote off those they think should be removed from the competition. We refer the reader to \url{https://github.com/mariamwalaa/CON-CN} for access to the data, models, and further analysis.

The algorithm computes first-order and second-order CON scores in a competition network where nodes represent actors and directed edges represent one actor defeating another in a specific round of the competition. The algorithm iterates over competitions and rounds and independently computes the CON score for each actor in each round. The \emph{Node Ranking Model} (or NRM) for Survivor is as follows. The algorithm for Chess.com and Dota 2 is almost identical, except without the for loop in line 1.

\smallskip

\renewcommand{\thealgocf}{} % Remove numbering from Algorithm

\begin{algorithm}[H]
    \SetAlgoLined
    \ForEach{competition in competitions}{
        \ForEach{round in competition}{
            Generate a graph of the current round\\
            Construct adjacency matrix for actors\\
            Square adjacency matrix to consider second-order neighbors\\
            Calculate 1st and 2nd order CON scores for each actor\\
            Store CON scores in a results list
        }
    }
    \Return results
    \caption*{Dynamic CON score in Survivor.}
\end{algorithm}

\begin{enumerate}
    \item \emph{Network Construction}: Competitions that have occurred have associated data posted online. Once data for a given competition and its associated rounds have been collected, a network is constructed by identifying teams, players, and the winners of each game. Four main variables are of interest: competition, round, game/match, players/teams, and time-step. If multiple competitions occur, then the competition-specific networks are amalgamated into one large network.
        
    \item \emph{Feature Generation}: Once a network is constructed, centrality measures are computed for every node at each time step $1$ through $k$, and a feature matrix is generated with $n$ rows representing teams/players and $m$ columns representing centrality features. 
    
    \item \emph{Ground Truth Labels}: To train the models, a classification of nodes based on ground truth data must be performed. We use the current rankings of players as well as known outcomes to create these classes. The numeric rankings are categorized into three quantile groups: \emph{the bottom} 10\%, \emph{the top} 10\%, and the \emph{middle range} consisting of the remaining 80\%.
        
    \item \emph{Model Training}: The final step involves first splitting the feature data into training and testing sets, then training machine learning models on data points associated with the first 80\% of time steps and testing them on the data points associated with the remaining 20\% of time steps.
\end{enumerate}

We use supervised machine learning models built for classification purposes to predict node importance. These models are well-suited for handling high-dimensional tabular data, which can be derived from network features by transforming them into a structured format. This transformation enables us to leverage the information encoded in networks, allowing us to take advantage of classification models. A sample of this transformed data is presented in Table~1.

\begin{table}[h!]
    \centering
    \begin{tabular}{|c|c|c|c|c|c|}
    \hline
    \textbf{Team} & \textbf{ CON } & \textbf{ PageRank } & \textbf{ Closeness } & \textbf{ In-Degree } & \textbf{ Out-Degree } \\
    \hline
    8629005.0 & 1319.0 & 0.008925 & 0.027778 & 6.0 & 7.0 \\
    8629317.0 & 1216.0 & 0.009633 & 0.025641 & 5.0 & 6.0 \\
    9426115.0 & 1212.0 & 0.006819 & 0.025641 & 5.0 & 6.0 \\
    9425660.0 & 1204.0 & 0.004711 & 0.023810 & 5.0 & 7.0 \\
    9425656.0 & 1176.0 & 0.008009 & 0.025641 & 5.0 & 7.0 \\
    \hline
    \end{tabular}
    \caption{Dota 2 data sample from Week 39, sorted by CON score. The Dota 2 team corresponding to ID 8629005.0 had a CON score of 1319, a PageRank score of 0.008925, a closeness score of 0.027778, in-degree of 6, and out-degree of 7.  Table only shows metrics for one round of competitions.}
\end{table}

We start with a decision tree, a simple model that splits data based on a series of hierarchical feature conditions to classify observations. However, as decision trees are prone to overfitting, we also consider a random forest model, which is a robust ensemble of decision trees. To further enhance predictive performance, we also explore gradient boosting and XGBoost \cite{xgboost}, which are better at penalizing overfitting by building trees sequentially while optimizing for errors. Finally, we consider a support vector machine, which falls under a different class of machine learning algorithms and classifies data by finding the best hyperplane; that is, the largest separation between classes.

Table~2 provides details about the size of the networks and how long it takes to run the NRM on each of the datasets. The difference in the size of the edge sets is reflected in the time in seconds it takes to run the model. The performance metrics of the machine learning models are in Table~3. We consider the F1 score rather than the accuracy of each dataset since the classes are imbalanced. We see that the Dota 2 network has the highest accuracy at 0.845 while the Survivor network has the highest F1 score at 0.788.

\begin{table}[h!]
    \centering
    \renewcommand{\arraystretch}{1.2} 
    \begin{tabular}{|l|ccc|}
        \hline
        \textbf{Metric} & \textbf{Survivor} & \textbf{Chess.com} & \textbf{Dota 2} \\ \hline
        \# Nodes        & 806               & 933                & 493            \\ 
        \# Edges        & 3,662             & 16,571             & 2,413          \\ 
        \# Rounds       & 12                & 18                 & 8              \\ 
        \# Competitions & 46                & 1                  & 1              \\ 
        \# Labels       & 152/455/152       & 87/690/86          & 50/393/50      \\ 
        Connected       & No                & Yes                & No             \\ 
        \# WCC          & 46                & 1                  & 39             \\ 
        \# SCC          & 90                & 152                & 199            \\ 
        Sparsity        & 0.0064            & 0.0214             & 0.0081         \\ 
        Diameter        & 3                 & 4                  & 10             \\ 
        Runtime         & 1.5s          & 635s               & 22s            \\ \hline
        \end{tabular}
    \caption{Graph descriptions for Survivor, Chess.com, and Dota 2 datasets. The Dota 2 network has the highest diameter of 10 as well as the lowest sparsity of 0.0064, indicating that it is the most spread-out, dense network, while the Survivor network has the highest number of weakly connected components (or WCCs), a total of 46.}
\end{table}

\begin{table}[h!]
    \centering
    \renewcommand{\arraystretch}{1.2} 
    \begin{tabular}{|l|cccc|cccc|cccc|}
        \hline
        & \multicolumn{4}{c|}{\textbf{Survivor}} 
        & \multicolumn{4}{c|}{\textbf{Chess.com}} 
        & \multicolumn{4}{c|}{\textbf{Dota 2}} \\ \hline
        \scriptsize{\textbf{Model}} 
        & \textbf{Acc.} & \textbf{Prec.} & \textbf{Rec.} & \textbf{F1} 
        & \textbf{Acc.} & \textbf{Prec.} & \textbf{Rec.} & \textbf{F1} 
        & \textbf{Acc.} & \textbf{Prec.} & \textbf{Rec.} & \textbf{F1} \\ \hline
        SVM 
        & \textbf{0.816} & \textbf{0.780} & \textbf{0.807} & \textbf{0.788} 
        & 0.799 & 0.266 & 0.333 & 0.296 
        & 0.608 & 0.570 & 0.662 & 0.549 \\ 
        RF 
        & 0.750 & 0.705 & 0.697 & 0.700 
        & \textbf{0.811} & 0.472 & \textbf{0.450} & 0.452 
        & 0.770 & 0.645 & 0.594 & 0.606 \\ 
        XGB 
        & 0.750 & 0.713 & 0.695 & 0.701 
        & 0.803 & \textbf{0.602} & 0.436 & 0.461 
        & 0.791 & 0.628 & 0.602 & 0.612 \\ 
        GB 
        & 0.728 & 0.685 & 0.659 & 0.666 
        & 0.788 & 0.423 & 0.429 & 0.422 
        & \textbf{0.845} & \textbf{0.714} & \textbf{0.663} & \textbf{0.686} \\ 
        DT 
        & 0.662 & 0.607 & 0.566 & 0.578 
        & 0.699 & 0.446 & 0.493 & \textbf{0.462} 
        & 0.777 & 0.615 & 0.635 & 0.620 \\ \hline
    \end{tabular}
    \caption{Performance metrics for Survivor, Chess.com, and Dota 2 datasets across various models: support vector machine, random forest, XGBoost, gradient boosting, and decision trees. The highest score for each metric is highlighted in bold.}
\end{table}

To determine which centrality features are most important for the top-performing machine learning model in each dataset, we use tree-based feature importance, which measures the \emph{mean decrease in impurity} (or MDI) as
\[
\text{MDI}(f_i) = \sum_{t \in T_i} p(t) \Delta I_t,
\]
where $T_i$ is the set of all tree nodes in the forest where $f_i$ is used for splitting, $p(t)$ is the proportion of samples reaching node $t$ (node $t$'s weight in the tree), and $\Delta I_t$ is the impurity decrease achieved at node $t$. The larger the MDI, the more important a feature is deemed to the model. As shown in Figure~3, the feature importance of the CON score exceeds that of the remaining centrality measures in the Dota 2 dataset. Specifically, we see that there is an uptick in importance corresponding to week 39, which suggests that CON scores of teams who played in week 39 were crucial for predicting the class of teams. 

\begin{figure}[h!]
    \centering
    \begin{minipage}{0.8\textwidth}
        \includegraphics[width=\textwidth]{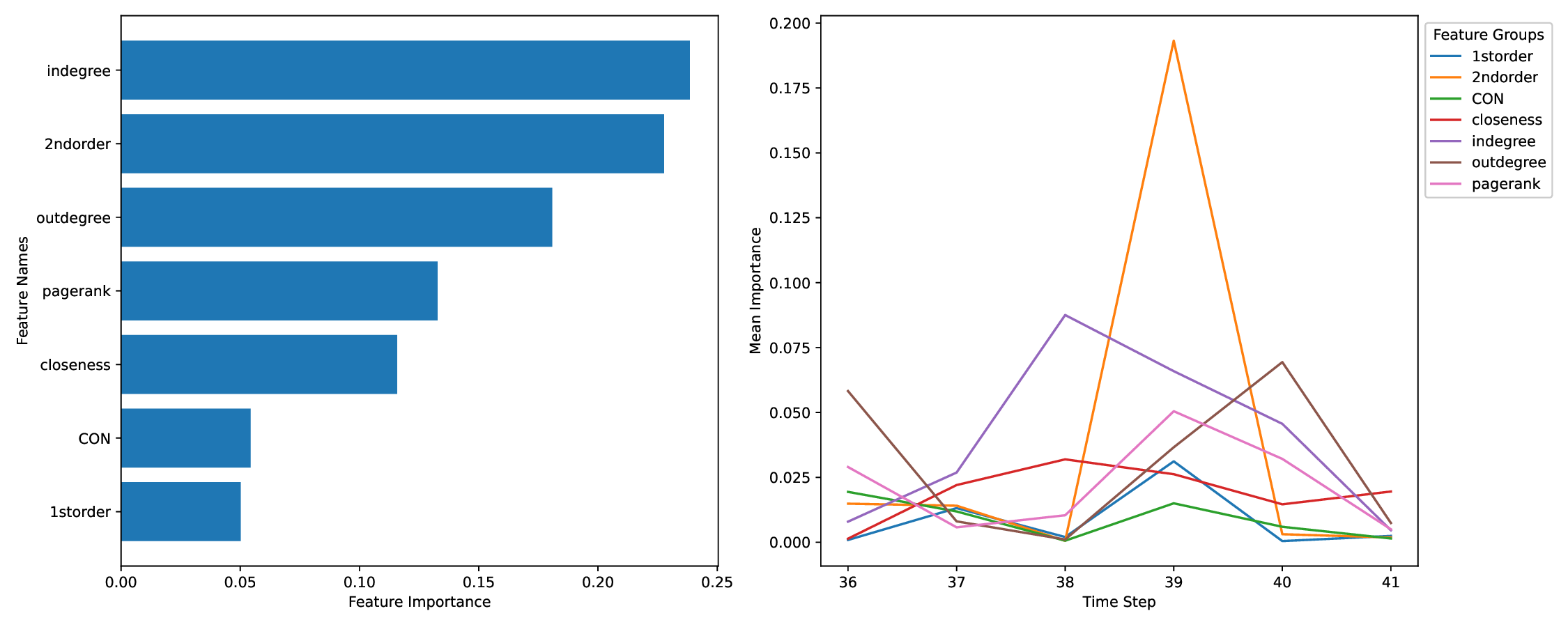}
        \caption{On the left, a bar chart of Dota 2 mean feature importance (y-axis) versus time-step (x-axis) shows that the CON score has a higher mean feature importance than other centrality measures across all time-steps. On the right, a line chart shows mean importance at each time-step, with a jump in CON mean importance at week 39.}
    \end{minipage}
\end{figure}

To assess the relationship between centrality metrics and ground truth labels (such as team rankings or player performance), we rank-order the centrality measures and display them as a line chart. A matrix of line charts for the Chess.com dataset is shown in Figure~4. First, we apply a re-scaling technique known as \emph{unity-based normalization}, as described in \cite{bonato2022}. Given data points \(X_1, X_2, \ldots, X_n\), with minimum value \(X_{\text{min}}\) and maximum value \(X_{\text{max}}\) where \(1 \leq i \leq n\), the normalized value \(X_{i,\text{norm}}\) is calculated as 
\[ 
X_{i,\text{norm}} = \frac{X_i - X_{\text{min}}}{X_{\text{max}} - X_{\text{min}}}. 
\] 
This scaling method ensures that all normalized values \(X_{i,\text{norm}}\) lie within the range \([0, 1]\). For consistency, we apply this normalization method to all centrality measures. In addition to scaling, a 50-point moving average is applied to smooth the data and better visualize the trend. 

In Table~4, Spearman’s coefficient is used to measure the correlation between the predicted rankings from centrality-based measures and actual ground truth labels. The results show strong correlations, suggesting that centrality features can effectively approximate the true influence of nodes in competition networks. As shown in Figure~4 and Table~4, there is a high correlation between the CON score and common centrality measures, which supports the hypothesis that the CON score can be used as a centrality measure.

\begin{figure}[h!]
    \includegraphics[width=1.0\textwidth]{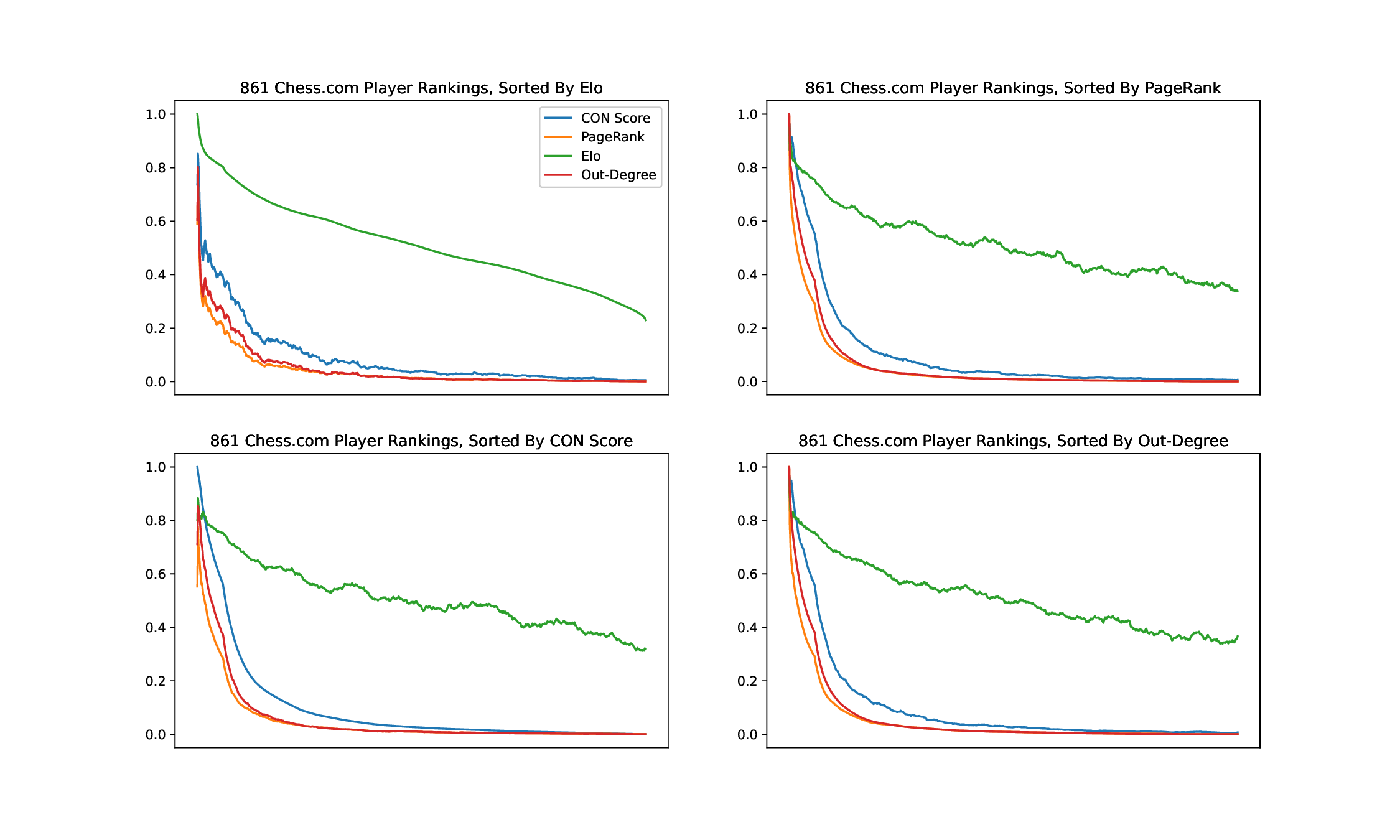}
    \caption{A 2-by-2 matrix of line charts shows four measures (CON, PageRank, Elo, and Out-Degree) for 861 Chess.com players, each sorted by one measure. A negative exponential correlation exists between all centralities and the sorted measure of choice.}
\end{figure}

\begin{table}[h!]
    \centering
    \renewcommand{\arraystretch}{1.2} 
    \begin{tabular}{|l|c|c|}
    \hline
    \textbf{ Metric } & \textbf{ Correlation } & \textbf{ P-Value } \\ \hline
    Out-Degree      & 0.736                       & \(6 \times 10^{-148}\) \\ 
    CON Score       & 0.707                       & \(2 \times 10^{-131}\) \\ 
    PageRank        & 0.701                       & \(2 \times 10^{-128}\) \\ 
    Betweenness     & 0.664                       & \(2 \times 10^{-110}\) \\ 
    Closeness       & 0.309                       & \(2 \times 10^{-20}\)  \\ \hline
    \end{tabular}
    \caption{Spearman correlation coefficients (sorted in descending order) between centrality measures and Chess.com rankings show that out-degree has the highest correlation to Chess.com rankings, while closeness has the lowest correlation to Chess.com known rankings.}
\end{table}

\begin{figure}[h!]
    \centering
    \begin{minipage}{0.32\textwidth}
        \centering
        \includegraphics[width=\textwidth]{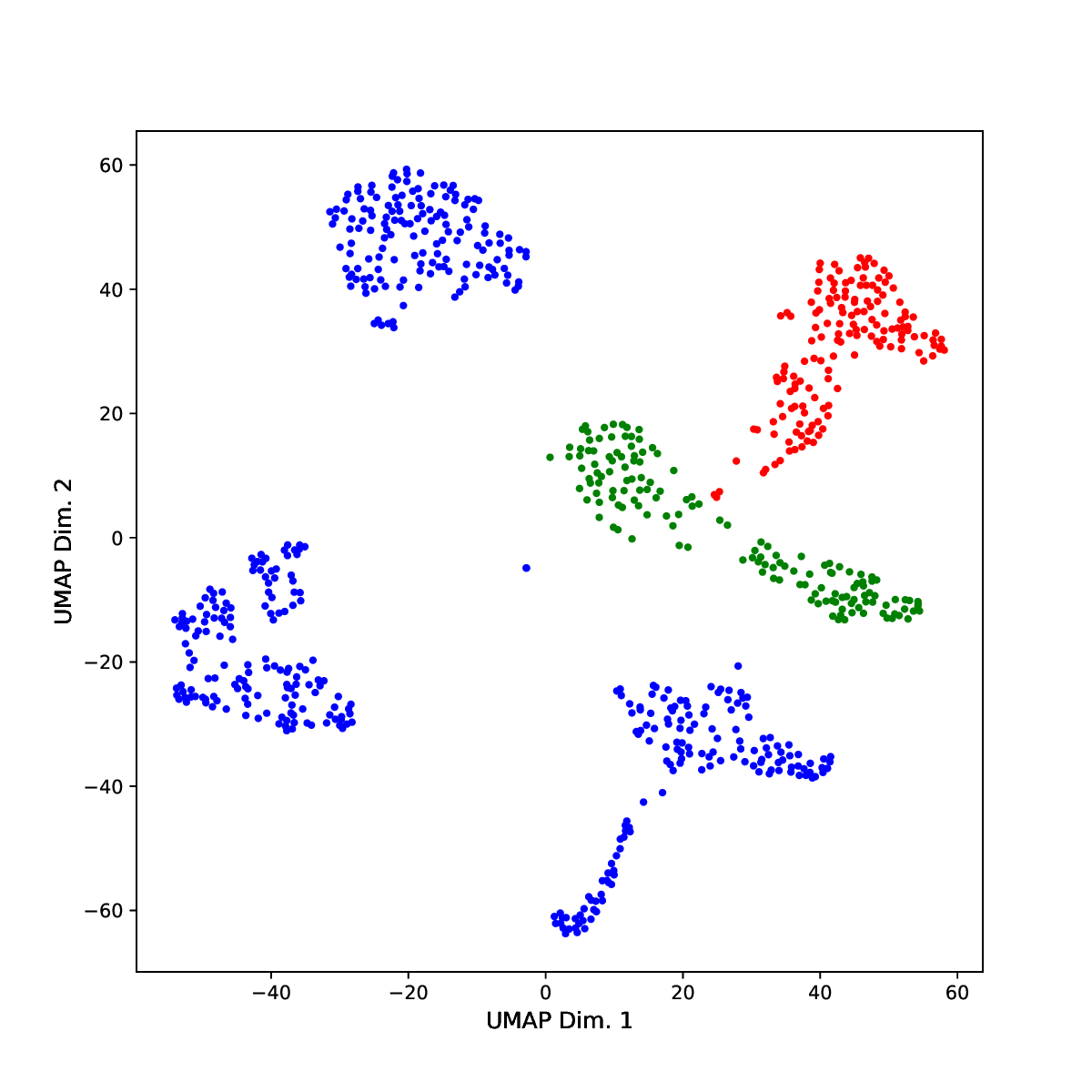}
        \caption*{(a) Survivor}
    \end{minipage}
    \hfill
    \begin{minipage}{0.32\textwidth}
        \centering
        \includegraphics[width=\textwidth]{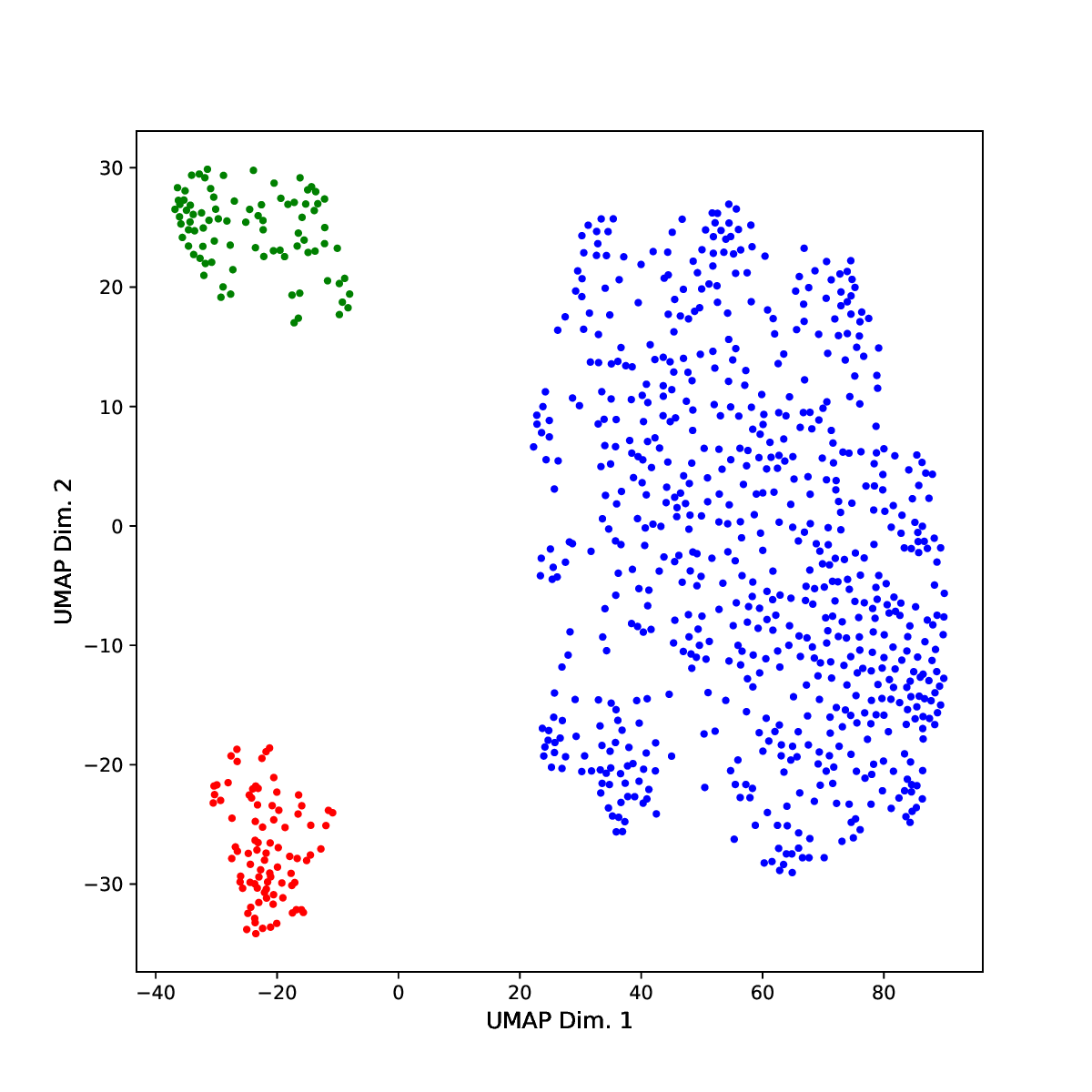}
        \caption*{(b) Chess}
    \end{minipage}
    \hfill
    \begin{minipage}{0.32\textwidth}
        \centering
        \includegraphics[width=\textwidth]{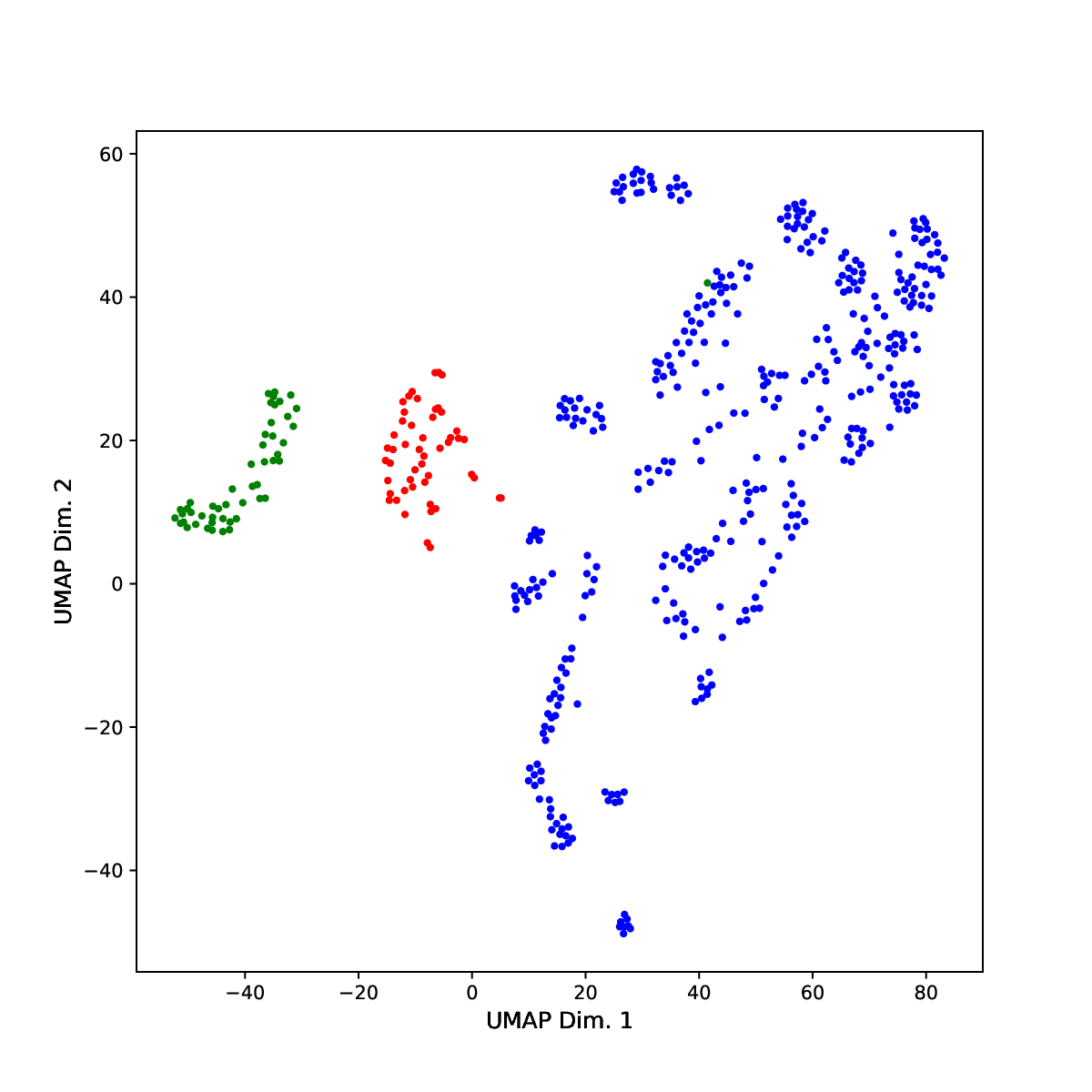}
        \caption*{(c) Dota 2}
    \end{minipage}
    \caption{Plots of the 2-dimensional supervised UMAP embeddings for each actor in the competition networks, where red is the top 10\%, green is the bottom 10\%, and blue is the middle 10\%-90\%. (a) shows that Survivor castaways in the middle range are dispersed into three main groups, whereas the top and bottom ranges are more clustered together. (b) shows that Chess.com players in the middle range are more evenly distributed, while the top and bottom ranges are distant from each other. (c) shows that the Dota 2 middle range has many smaller clusters, while the top and bottom ranges are close. The maps show that there is a clear separation between classes with localized clustering.}
\end{figure}
To further validate our findings, we generated a supervised \emph{Uniform Manifold Approximation and Projection} (or UMAP) \cite{mcinnes2018} embedding of the model features to determine whether these features properly distinguish between the three classes of nodes. See Figure~5. The approach also shows a clear separation between the three classes and localized clustering within the middle class. For example, the middle class in the Survivor dataset consists of three distant clusters, while the Dota 2 middle class has smaller, more dispersed clusters. This aligns with expectations, as the middle class contains the majority of the observations and is likely to exhibit greater variation.

\section{Discussion and Future Work}

We developed a method to dynamically compute the CON score and other common centrality measures for competition networks. This approach enabled the creation of a feature set that accurately predicts the class of a node based on ground truth data. Our analysis confirmed that the CON score effectively seriated nodes, as evidenced by a high Spearman coefficient and strong visual correlation between metrics at the node level.

While we demonstrated aggregate feature importance to establish the strength of the CON score relative to other centrality measures, future work could focus on node-level analysis of feature importance to gain deeper insights. Model performance may also benefit from improved ground truth labels. For instance, in the Dota 2 datasets, team rankings are more closely aligned with the timeline of matches, and additional context about the data collection process could enhance predictions. Beyond games, adversarial networks in fields like biology, such as food webs and dominance hierarchies, offer promising avenues for further exploration.

Future research could apply this algorithm to various adversarial networks and random graph models, including Erdős–Rényi graphs, the preferential attachment model, or random geometric graphs, to assess its broader applicability. Questions about the CON score's nature, such as whether it is a local or global metric, remain to be addressed. Additionally, generalizing the CON score to the $k$-th order, where $k$ is the diameter or another meaningful parameter, could provide novel insights into network structure and dynamics. By addressing these challenges, we aim to expand the utility of the CON score and contribute to a deeper understanding of complex networks across disciplines.

\end{document}